\documentclass[12pt]{JHEP3}
\usepackage{amssymb,amsmath,amsthm}
\usepackage{mathrsfs} %script fonts
\usepackage{epsfig,multicol,bbm}
\def\TODAY{15 April 2010; 21 April 2010; 22 July 2010}
\title{Semi-analytic results for quasi-normal frequencies}
\author{{Jozef Skakala} and {Matt Visser}\\
School of Mathematics, Statistics, and Operations Research, \\
Victoria University of Wellington, 
Wellington, New Zealand\\
E-mail: \email{jozef.skakala@msor.vuw.ac.nz, matt.visser@msor.vuw.ac.nz}
}
\received{\TODAY;  \LaTeX-ed \today}                                           
\abstract{The last decade has seen considerable interest in the quasi-normal frequencies [QNFs] of black holes (and even wormholes), both asymptotically flat and with cosmological horizons.  There is wide agreement that the QNFs are often of the form $\omega_n =  \hbox{(offset)} + i n \; \hbox{(gap)}$,
though some authors have encountered situations where this behaviour seems to fail.  To get a better understanding of the general situation we consider a semi-analytic model based on a \emph{piecewise} Eckart (P\"oschl--Teller) potential, allowing for different heights and different rates of exponential falloff in the two asymptotic directions. This model is sufficiently general to capture and display key features of the black hole QNFs while simultaneously being analytically tractable, at least for asymptotically large imaginary  parts of the QNFs. 

We shall derive an appropriate ``quantization condition'' for the asymptotic QNFs, and extract as much analytic information as possible. In particular, we shall explicitly verify that the  $ \hbox{(offset)} + i n \; \hbox{(gap)}$ behaviour is \emph{common but not universal}, with this behaviour failing unless the ratio of rates of exponential falloff on the two sides of the potential is a rational number. (This is ``common but not universal'' in the sense that the rational numbers are dense in the reals.) We argue that this behaviour is likely to persist for black holes with cosmological horizons. 
}


\begin{document}
%------------------------------------------------------------------------------------------------------------------------------------------

%\clearpage
%------------------------------------------------------------------------------------------------------------------------------------------
% very standard definitions
%------------------------------------------------------------------------------------------------------------------------------------------
\def\d{{\mathrm{d}}}
\newcommand{\scri}{\mathscr{I}}
\newcommand{\sun}{\ensuremath{\odot}}
\def\J{{\mathscr{J}}}
\def\sech{{\mathrm{sech}}}
\def\T{{\mathcal{T}}}
\def\Im{{\mathrm{Im}}}
\def\Re{{\mathrm{Re}}}
\tableofcontents
%\clearpage
%------------------------------------------------------------------------------------------------------------------------------------------
%-----------------------------------------------------------------------------------------------------------------------------------------
\section{Introduction}
%-----------------------------------------------------------------------------------------------------------------------------------------

Black hole quasi-normal frequencies [QNFs] have been extensively studied over the past several decades, both for (potential) observational reasons, and for a number of highly technical theoretical reasons~\cite{Chandrasekhar, Kokkotas, Nollert, Berti, Dreyer, Natario}. A key point is that the Regge--Wheeler equation, and the Zerelli equation,  can (when expressed in terms of the tortoise $r_*$ coordinate) be written in the form
\begin{equation}
{d^2 \psi\over d r_*^2} =   [ \omega^2 - V(r_*) ] \; \psi.
\end{equation}
The QNFs arising from this differential equation are in principle observable in astrophysical black hole oscillations and ring-down phenomena. At an abstract theoretical level it has been observed that the QNFs associated with this Regge--Wheeler equation are very often of the form
\begin{equation}
\omega_n =  \hbox{(offset)} + i n \; \hbox{(gap)}.
\end{equation}
Such behaviour was first noted when using the Eckart potential~\cite{Eckart} (P\"oschl--Teller potential~\cite{Poeschl, Morse})\footnote{The nomenclature in this field is quite confusing, with the phrase ``P\"oschl--Teller potential'' being ascribed somewhat erratically to a number of different quantities. In this article we will always be referring to some variant of the $\sech^2(\cdot)$ potential, and will adopt the historically more accurate nomenclature of calling this (a special case of) the Eckart potential~\cite{Eckart}. (See~\cite{Boonserm, Boonserm2} for some general discussion of this historical point.)}  to estimate the lowest lying QNFs~\cite{Ferrari, Iyer, Guinn1}, but since then attention has focussed more on the higher lying QNFs as the mode number $n \to \infty$~\cite{Dreyer, Guinn2, Konoplya}. For $n\to\infty$ the ``gap" is simply related to the surface gravity of the black hole, at least in the asymptotically flat case where there is only a single horizon. The situation where both event and cosmological horizons are present is more subtle, and we shall seek a better semi-analytic understanding of this case by investigating a simplified model based on a \emph{piecewise}  Eckart potential (P\"oschl--Teller potential).  This model has the virtues of being analytically tractable (at least for high lying QNFs as the mode number $n \to \infty$) while being general enough to serve as a plausible model capturing the essence of the physics for a two-horizon black hole.

We shall derive an appropriate ``quantization condition'' for the asymptotic QNFs, and from this condition extract as much analytic information as possible. In particular, we explicitly verify that the  $ \hbox{(offset)} + i n \; \hbox{(gap)}$ behaviour is \emph{common but not universal}, with this behaviour failing unless the ratio of rates of exponential falloff on the two sides of the potential is a rational number.  We argue that this behaviour is likely to persist for real asymptotically de Sitter black holes with both event and cosmological horizons. 

This  $ \hbox{(offset)} + i n \; \hbox{(gap)}$ behaviour  is ``common but not universal'' in the sense that the rational numbers are dense in the reals.  Furthermore, if one resorts to numerical techniques, then since the floating point numbers are rational approximations to real numbers,   the $ \hbox{(offset)} + i n \; \hbox{(gap)}$ behaviour is in fact \emph{universal} within the context of floating point arithmetic, and all numerical experiments along these lines should be interpreted with this point in mind.

%-----------------------------------------------------------------------------------------------------------------------------------------
\section{Black hole QNFs}
%-----------------------------------------------------------------------------------------------------------------------------------------

For the specific case of a Schwarzschild black hole the tortoise coordinate is given by
\begin{equation}
{dr\over dr_*} = 1-{2m\over r}; \qquad \qquad r_*(r) = r + 2m\ln\left[{r-2m\over2m}\right];
\end{equation}
and the Regge--Wheeler potential is
\begin{equation}
V(r_*) = \left(1-{2m\over r}\right) \left[ {\ell(\ell+1)\over r^2} + {2m(1-s^2)\over r^3} \right].
\end{equation}
Here $s$ is the spin of the particle and $\ell$ is the angular momentum of the specific wave mode under consideration, with $\ell\geq s$. 
As $r_*\to -\infty$ we have $r\to 2m$ and
\begin{equation}
V(r_*)\to  \exp\left( {r_*-2m\over 2m} \right) \; {\ell(\ell+1) +(1-s^2) \over (2m)^2}  = V_0 \;  \exp( 2 \kappa r_* ),
\end{equation}
where $\kappa$ is the black hole surface gravity. This specific behaviour in terms of the surface gravity generalizes beyond the Schwarzschild black hole and for an arbitrary black hole in an asymptotically flat spacetime one has
\begin{equation}
V(r_*) \to \left\{ \begin{array}{ll}
\vphantom{\Big|}
 V_{0-} \; \exp( - 2\kappa  |r_*|  ),   & r_* \to - \infty;
 \\
 \vphantom{\Big|}
  V_{0+}  \; (2m)^2/ r^2,  & r_* \to + \infty.
\end{array} \right.
\end{equation}
This is often approximated (rather brutally) by an Eckart potential (P\"oschl--Teller potential) 
\begin{equation}
V(r_*) = {V_0 \over \cosh^2(\kappa r_*)},
\end{equation}
or more generally by using two adjustable parameters $V_0$ and $b$
\begin{equation}
V(r_*) = {V_0 \over \cosh^2( r_*/b)}.
\end{equation}
If one is interested in probing the lowest-lying QNFs one choses $V_0$ and $b$ to match the peak and peak curvature of the Regge--Wheeler potential~\cite{Ferrari, Iyer, Guinn1, Guinn2, Konoplya}. If instead one is interested in the higher QNFs one picks $b=1/\kappa$ (and in this case chooses $V_0$ in some more or less arbitrary manner)~\cite{Medved1, Medved2, Padmanabhan, Choudhury}.
Though the approximation is brutal, the Eckart potential has the great benefit of being analytically soluble, and the QNFs are known exactly. (See for example~\cite{Beyer}, or the discussion in~\cite{Morse}. See also~\cite{Boonserm2}.) Taking 
\begin{equation}
\alpha = \left\{  \begin{array}{lcl}
\sqrt{{1\over4} - V_0 b^2 } & \hbox{ for }  & V_0 b^2 < 1/4,\\
\\
i \sqrt{V_0 b^2 - {1\over4}  } & \hbox{ for }  & V_0 b^2 > 1/4,\\
\end{array}\right.
\end{equation}
the exact QNFs are
\begin{equation}
\omega_{n,\pm} =  {i(n+{1\over2} \pm \alpha)\over b};  \qquad (n\in \{1,2,3,\dots\}),
\end{equation}
which is of the general form $\omega_n =  \hbox{(offset)} + i n \; \hbox{(gap)}$ with two families of QNFs described by the specific coefficients 
\begin{equation}
 \hbox{(offset)}  = {i({1\over2} \pm \alpha)\over b};  \qquad \qquad \hbox{(gap)} = {1\over b}. 
\end{equation}
Similar asymptotic behaviour for more realistic approximations to the Regge--Wheeler potential  has been verified by a number of independent approaches. For Schwarzschild black holes, and more generally single-horizon black holes, it has been demonstrated that:
\begin{itemize}
\item
WKB inspired approximations~\cite{Ferrari, Iyer, Guinn1, Guinn2, Konoplya};
\item 
phase-amplitude methods~\cite{Andersson1, Andersson2};
\item
continued fraction approximations for the QNFs~\cite{Leaver1, Leaver2, Leaver3};
\item
monodromy techniques~\cite{Motl1, Motl2, Natario, Shanka1, Shanka2};
\item
Born approximations~\cite{Medved1, Medved2, Padmanabhan, Choudhury};
\end{itemize}
all agree on this general  $\omega_n =  \hbox{(offset)} + i n \; \hbox{(gap)}$ form for the QNFs with the surface gravity $\kappa \leftrightarrow 1/b$ governing the ``gap'' between  the high-lying modes. For Reissner--Nordstr\"om, Kerr, Kerr--Newman, Sen [Reissner--Nordstr\"om-dilaton],  black holes the situation is somewhat messier and sometimes still controversial, with results often being dimension dependent, and with results often (somewhat unexpectedly) depending on the properties of the inner horizon~\cite{Chan1, Chan2, Horowitz, Wang1, Wang2, Wang3, Cardoso, Wang4, Wang5, Wang6}. (See in particular~\cite{Cardoso} for a recent review.) For example, even for the Reissner--Nordstr\"om black hole the $Q/M\to 0$ limit does not seem to reproduce the QNFs of the Schwarzschild black hole, and situations where $\sqrt{1-Q^2/M^2}$ is a rational number seem to be special. Though the techniques we adopt below are not directly applicable to the Reissner--Nordstr\"om black hole it is perhaps interesting to note that the special situations where  $\sqrt{1-Q^2/M^2}$ is a rational number correspond to $\kappa_+/\kappa_-$, the ratio of surface gravities at the inner and outer horizons, being the square of a rational number.

If one turns to asymptotically de Sitter black holes (or more generally any two horizon system where the region of interest is bounded by the two horizons --- an event horizon and a cosmological horizon) the situation is considerably more ambiguous --- the Regge--Wheeler potential and the Zerelli potential have the asymptotic behaviour
\begin{equation}
V(r_*) \to \left\{ \begin{array}{ll}
\vphantom{\Big|}
 V_{0-} \; \exp( - 2\kappa_-  |r_*|  ),   &\qquad  r_* \to - \infty;
 \\
 \vphantom{\Big|}
  V_{0+}  \;  \exp( - 2\kappa_+  |r_*|  ),  &\qquad  r_* \to + \infty;
\end{array} \right.
\end{equation}
where the two surface gravities are now (in general) distinct. It is this situation that we will model using a \emph{piecewise}  Eckart potential (P\"oschl--Teller potential).  A considerable amount of analytic information can be extracted from this model, information which in the concluding discussion we shall attempt to relate back to ``realistic'' black hole physics.

%-----------------------------------------------------------------------------------------------------------------------------------------
\section{Piecewise Eckart (P\"oschl--Teller) potential}
%-----------------------------------------------------------------------------------------------------------------------------------------

%-----------------------------------------------------------------------------------------------------------------------------------------
\subsection{Potential}
%-----------------------------------------------------------------------------------------------------------------------------------------

The model  we are interested in investigating is
\begin{equation}
-\psi''(x) + V(x) \; \psi(x) = 0,
\end{equation}
with 
\begin{equation}
V(x) = \left\{  \begin{array}{lcl}
{V_{0-} \; \sech^2(x/b_-)}  & \hbox{ for }  & x < 0;\\
\\
V_{0+} \; \sech^2(x/b_+)  & \hbox{ for }  & x > 0. \\
\end{array}\right.
\end{equation}
We will allow a discontinuity in the potential at $x=0$. The standard case that is usually dealt with is for
\begin{equation}
V_{0-} = V_{0+} = V_0;   
\qquad\qquad 
b_-=b_+ = b; 
\qquad\qquad
V(x) = {V_{0} \over \cosh^2(x/b)}.
\end{equation}
A related model where $V_{0-} = V_{0+} = V_0$ but $b_+\neq b_-$ has been explored by Suneeta~\cite{Suneeta}, but our current model is more general, and we will take the analysis much further. 

%-----------------------------------------------------------------------------------------------------------------------------------------
\subsection{Wavefunction}
%-----------------------------------------------------------------------------------------------------------------------------------------

We start by imposing quasi-normal boundary conditions (outgoing radiation boundary conditions)
\begin{equation}
\psi_+(x\to+\infty) \to e^{-i\omega x}; \qquad \psi_-(x\to-\infty) \to e^{+i\omega x}.
\end{equation}
On each half line ($x<0$, and $x>0$) the exact wavefunction  (see especially page 405 of the article by Beyer~\cite{Beyer}) is:
\begin{equation}
\psi_\pm(x) = e^{\mp i\omega x} \; {} _2F_1\left({1\over2}+\alpha_\pm,{1\over2}-\alpha_\pm,1+ib_\pm \omega, {1\over 1+ e^{\pm 2x/b_\pm}} \right),
\end{equation}
where
\begin{equation}
\alpha = \left\{  \begin{array}{lcl}
\sqrt{{1\over4} - V_0 b^2 } & \hbox{ for }  & V_0 b^2 < 1/4;\\
\\
i \sqrt{V_0 b^2 - {1\over4}  } & \hbox{ for }  & V_0 b^2 > 1/4.\\
\end{array}\right.
\end{equation}
``All'' we need to do is to appropriately match these wavefunctions at the origin.
%

%-----------------------------------------------------------------------------------------------------------------------------------------
\subsection{Junction condition}
%-----------------------------------------------------------------------------------------------------------------------------------------

The key step in matching these two wavefunctions at $x=0$ is to calculate the logarithmic derivative. Using the variable $z = 1/(1+ e^{\pm 2x/b_\pm})$, note that $x=0$ maps into $z=1/2$. Then using the Leibnitz rule and the chain rule one has:
\begin{equation}
{\psi_\pm'(0)\over\psi_\pm(0)} = 
\mp i \omega \mp    {1\over 2 b_\pm} \left. {\d \ln \left\{ _2F_1\left({1\over2}+\alpha_\pm,{1\over2}-\alpha_\pm,1+ib_\pm \omega,z\right) \right\} \over \d z}\right|_{z=1/2.} 
\end{equation}
Invoking the differential identity (\ref{E:differential}) in appendix \ref{A:hyper}, we see
\begin{equation}
{\psi_\pm'(0)\over\psi_\pm(0)} = \mp i \omega \; 
\left.
{_2 F_1\left({1\over2}+\alpha_\pm,{1\over2}-\alpha_\pm,ib_\pm \omega,z\right)
\over
_2F_1\left({1\over2}+\alpha_\pm,{1\over2}-\alpha_\pm,1+ib_\pm \omega,z\right)}
\right|_{z=1/2.} 
\end{equation}
(In special situations where $\psi_\pm(0)$ might accidentally equal zero one might need to perform a special case analysis. The generic situation is $\psi_\pm(0)\neq 0$, and will prove sufficient for almost everything we need to calculate.) 
Now using Bailey's theorem (\ref{E:bailey}) to evaluate the hypergeometric functions at $z\to{1\over2}$ we have the exact result
\begin{equation}
{\psi_\pm'(0)\over\psi_\pm(0)} 
= \mp {2\over b_\pm} \; {\Gamma({\alpha_\pm+i\omega b_\pm\over2} +{3\over4}) \Gamma({-\alpha_\pm+i\omega b_\pm\over2} +{3\over4}) \over 
\Gamma({\alpha_\pm+i\omega b_\pm \over2} + {1\over4}) \Gamma({-\alpha_\pm+i\omega b_\pm \over2} + {1\over4})}.
\end{equation}
Now if $\omega$ has a large positive imaginary part, then the Gamma function arguments above tend towards the negative real axis, a region where the Gamma function has many poles. This is computationally inconvenient, and to obtain a more tractable result it is extremely useful to use the reflection formula (\ref{E:reflection}) of appendix \ref{A:gamma} to derive
\begin{eqnarray}
 {\Gamma({\alpha_\pm+i\omega b_\pm\over2} +{3\over4}) \over \Gamma({\alpha_\pm+i\omega b_\pm \over2} + {1\over4}) } 
 &=&
 { \Gamma(1-{\alpha_\pm+i\omega b_\pm \over2} - {1\over4}) \sin(\pi [{\alpha_\pm+i\omega b_\pm \over2} + {1\over4}) ])
 \over
 \Gamma(1-{\alpha_\pm+i\omega b_\pm\over2} -{3\over4}) \sin(\pi[{\alpha_\pm+i\omega b_\pm\over2} +{3\over4}])}
 \nonumber
 \\
&=&  { \Gamma(-{\alpha_\pm+i\omega b_\pm \over2} + {3\over4}) \; \sin(\pi [{\alpha_\pm+i\omega b_\pm \over2} + {1\over4}) ])
 \over
 \Gamma(-{\alpha_\pm+i\omega b_\pm\over2} +{1\over4}) \sin(\pi[{\alpha_\pm+i\omega b_\pm\over2} +{1\over4}]+ {\pi\over2})}
\nonumber
 \\
&=&  { \Gamma(-{\alpha_\pm+i\omega b_\pm \over2} + {3\over4}) 
 \over
 \Gamma(-{\alpha_\pm+i\omega b_\pm\over2} +{1\over4})}  \times \tan\left(\pi\left[{\alpha_\pm+i\omega b_\pm\over2} +{1\over4}\right]\right).
 \end{eqnarray}
This leads to the exact result
\begin{eqnarray}
{\psi_\pm'(0)\over\psi_\pm(0)} &=&  \mp {2\over b_\pm}  \;
 { \Gamma({-\alpha_\pm-i\omega b_\pm \over2} + {3\over4})  \Gamma({\alpha_\pm-i\omega b_\pm \over2} + {3\over4}) 
 \over
 \Gamma({-\alpha_\pm-i\omega b_\pm\over2} +{1\over4})  \Gamma({\alpha_\pm-i\omega b_\pm\over2} +{1\over4})} 
 \nonumber
\\
&&
 \times
   \; \tan\left(\pi\left[{\alpha_\pm+i\omega b_\pm\over2} +{1\over4}\right]\right)  \; \tan\left(\pi\left[{-\alpha_\pm+i\omega b_\pm\over2} +{1\over4}\right]\right).
 \end{eqnarray}
If $\omega$ has a large positive imaginary part, then the Gamma function arguments above now tend towards the positive real axis, a region where the Gamma function is smoothly behaved --- all potential poles in the logarithmic derivative have been isolated in the trigonometric functions. 
We can also use one of the trigonometric  identities (\ref{E:trig1}) of appendix \ref{A:trig} to rewrite this as
  \begin{equation}
{\psi_\pm'(0)\over\psi_\pm(0)} =  \mp {2\over b_\pm}  \;
 { \Gamma({-\alpha_\pm-i\omega b_\pm \over2} + {3\over4})  \Gamma({\alpha_\pm-i\omega b_\pm \over2} + {3\over4}) 
 \over
 \Gamma({-\alpha_\pm-i\omega b_\pm\over2} +{1\over4})  \Gamma({\alpha_\pm-i\omega b_\pm\over2} +{1\over4})} 
 \times  {\cos(\pi\alpha_\pm) - \cos(\pi[i\omega b_\pm+1/2]) \over \cos(\pi\alpha_\pm) + \cos(\pi[i\omega b_\pm+1/2]) },
 \end{equation}
 which we can rewrite (still an \emph{exact} result) as
   \begin{equation}
{\psi_\pm'(0)\over\psi_\pm(0)} =  \mp {2\over b_\pm}  \;
 { \Gamma({-\alpha_\pm-i\omega b_\pm \over2} + {3\over4})  \Gamma({\alpha_\pm-i\omega b_\pm \over2} + {3\over4}) 
 \over
 \Gamma({-\alpha_\pm-i\omega b_\pm\over2} +{1\over4})  \Gamma({\alpha_\pm-i\omega b_\pm\over2} +{1\over4})} 
 \times  {\cos(\pi\alpha_\pm) + \sin(i\pi\omega b_\pm) \over \cos(\pi\alpha_\pm) - \sin(i\pi\omega b_\pm) }.
 \end{equation}
The exact junction condition we wish to apply at $x=0$ is
 \begin{equation}
{\psi_+'(0)\over\psi_+(0)}  = {\psi_-'(0)\over\psi_-(0)}, 
\end{equation} 
but the presence of the Gamma functions above makes this exact junction condition intractable. Fortunately, as long as we are primarily focussed on the highly damped QNFs ($\Im(\omega)\to\infty$) we can employ the Stirling approximation in the form (\ref{E:stirling}) indicated in appendix \ref{A:gamma} to deduce
\begin{eqnarray}
 { \Gamma({\pm\alpha_\pm-i\omega b_\pm \over2} + {3\over4})  
 \over
  \Gamma({\pm\alpha_\pm-i\omega b_\pm\over2} +{1\over4})  }
  &=&
  \sqrt{ {\pm\alpha_\pm-i\omega b_\pm\over2} +{1\over4} } \times  \left[1+ O\left({1\over \Im(\omega b_\pm)}\right)\right]
  \nonumber
  \\
  &=& \sqrt{ {\Im(\omega) b_\pm\over 2} }  \times \left[1+ O\left({1\over \Im(\omega b_\pm)}\right)\right].
  \end{eqnarray}
This allows us to deduce an \emph{approximate} junction condition for the asymptotic QNFs 
\begin{equation}
 {\cos(\pi\alpha_+) + \sin(i\pi\omega b_+) \over \cos(\pi\alpha_+) - \sin(i\pi\omega b_+) } = 
 -  {\cos(\pi\alpha_-) + \sin(i\pi\omega b_-) \over \cos(\pi\alpha_-) - \sin(i\pi\omega b_-) },
\end{equation}
which is accurate up to fractional corrections of order $O\left({1/ \Im(\omega b_\pm)}\right)$.
This is now an \emph{approximate} ``quantization condition'' for calculating the QNFs, which is asymptotically increasingly accurate for the highly-damped modes.

%-----------------------------------------------------------------------------------------------------------------------------------------
\subsection{QNF condition}
%-----------------------------------------------------------------------------------------------------------------------------------------

The asymptotic QNF condition above can, by cross multiplication and the use of trigonometric identities, be rewritten in any one of the four equivalent forms:
 \begin{equation}
 \label{E:qnf1}
\sin(-i\pi\omega b_+)\sin(-i\pi\omega b_-)  =  \cos(\pi\alpha_+)\cos(\pi\alpha_-);
 \end{equation}
  \begin{equation}
   \label{E:qnf2}
\sinh(\pi\omega b_+)\sinh(\pi\omega b_-)  =  - \cos(\pi\alpha_+)\cos(\pi\alpha_-);
 \end{equation}
\begin{equation}
 \label{E:qnf3}
\cos(-i\pi\omega[b_+ -b_-]) - \cos(-i\pi\omega[b_+ + b_-]) = 2  \;  \cos(\pi\alpha_+)\cos(\pi\alpha_-);
 \end{equation}
 \begin{equation}
 \label{E:qnf4}
\cosh(\pi\omega[b_+ -b_-]) - \cosh(\pi\omega[b_+ + b_-]) = 2  \;  \cos(\pi\alpha_+)\cos(\pi\alpha_-).
 \end{equation}
Which particular form one chooses to use is a matter of taste that depends on exactly what one is trying to establish. 
It is sometimes useful to split the asymptotic QNF condition into real and imaginary parts. To do so note
\begin{eqnarray}
\cos(A+iB) &=& \cos(A)\cos(iB) - \sin(A)\sin(iB)
\nonumber\\
& =& \cos(A) \cosh(B) - i \sin(A) \sinh(B),
\end{eqnarray}
so that 
\begin{eqnarray}
\cos(-i\pi\omega[b_+ -b_-]) &=& \cos(\Im(\omega)\pi[b_+ -b_-]) \cosh(\Re(\omega)\pi[b_+ -b_-])
\nonumber
\\
&&
 - i  \sin(\Im(\omega)\pi|b_+ -b_-|) \sinh(\Re(\omega)\pi|b_+ -b_-|).
\end{eqnarray}
Therefore the asymptotic QNF condition implies \emph{both}
\begin{eqnarray}
 \label{E:qnf:real}
&& 
\cos(\Im(\omega)\pi[b_+ -b_-]) \cosh(\Re(\omega)\pi[b_+ -b_-]) 
\nonumber
\\ &&  
\quad = \cos(\Im(\omega)\pi[b_+ +b_-]) \cosh(\Re(\omega)\pi[b_+ +b_-]) 
+ 2  \;  \cos(\pi\alpha_+)\cos(\pi\alpha_-),
\qquad
\end{eqnarray}
\emph{and}
\begin{eqnarray}
 \label{E:qnf:imaginary}
&&\sin(\Im(\omega)\pi|b_+ -b_-|) \sinh(\Re(\omega)\pi|b_+ -b_-|) 
\nonumber\\
&& \quad = \sin(\Im(\omega)\pi[b_+ +b_-]) \sinh(\Re(\omega)\pi[b_+ +b_-]).
\end{eqnarray}
We shall now seek to apply this QNF condition, in its many equivalent forms, to extract as much information as possible regarding the distribution of the QNFs.

%----------------------------------------------------------------------------------------------------------------------------------------
\section{Some general observations}\label{S:general}
%----------------------------------------------------------------------------------------------------------------------------------------

We shall start with some general observations regarding the QNFs.
\begin{enumerate}
\item Note that the $\alpha_\pm$ are either pure real or pure imaginary. 

\item Consequently $\cos(\pi\alpha_+)\cos(\pi\alpha_-)$ is always pure real $\in [-1, +\infty)$.

\item If  $\cos(\pi\alpha_+)\cos(\pi\alpha_-) > 0$, then there are no pure real QNFs.

\emph{Proof:} Consider equation (\ref{E:qnf2}) and note that under this condition the LHS is positive while the RHS is negative.

\item If  $\cos(\pi\alpha_+)\cos(\pi\alpha_-) < 0$, then there is a pure real QNF.

\emph{Proof:} Consider equation (\ref{E:qnf2}) and note that under this condition the RHS is positive. The LHS is positive and by continuity there will be a real root  $\omega \in (0,\infty)$.

\item If  $ \cos(\pi\alpha_+)\cos(\pi\alpha_-) > 1$, then there are no pure imaginary QNFs.

\emph{Proof:} Consider equation (\ref{E:qnf1}) and note that under this condition the LHS $\leq 1$ while the RHS $>1$.

\item There are infinitely many pure imaginary solutions to these asymptotic QNF conditions provided $ \cos(\pi\alpha_+)\cos(\pi\alpha_-)\leq Q(b_+,b_-) \leq 1$; that is, whenever  $ \cos(\pi\alpha_+)\cos(\pi\alpha_-)$  is ``sufficiently far'' below $1$. 

\emph{Proof:} Define
\begin{equation}
Q(b_+,b_-) =   \max_\omega\left\{  {\cos( |\omega|\pi[b_+ -b_-])   -  \cos( |\omega|\pi[b_+ +b_-])\over 2}  \right\} \leq 1.
\end{equation} 
Then by inspection equation (\ref{E:qnf3}) will have an infinite number of pure imaginary solutions as long as 
\begin{equation}
 \cos(\pi\alpha_+)\cos(\pi\alpha_-)\leq Q(b_+,b_-).
\end{equation}

\item For any purely imaginary $\omega$ there will be \emph{some} choice of $b_\pm$, $\alpha_\pm$ that makes this a solution of the asymptotic QNF condition.

\emph{Proof:} Consider the specific case
\begin{equation}
\omega =  {i (\alpha_+ + {1\over2})\over b_+} =  {i (\alpha_- + {1\over2})\over b_-},
\end{equation}
and note this satisfies the QNF condition but enforces only two constraints among the four unknowns $b_\pm$, $V_{0\pm}$. 

\end{enumerate}

%----------------------------------------------------------------------------------------------------------------------------------------
\section{Some approximate results}\label{S:approx}
%----------------------------------------------------------------------------------------------------------------------------------------

A number of approximate results can be extracted by looking at special regions of parameter space.

%-----------------------------------------------------------------------------------
\subsection{Case $b_- \approx b_+$}
%-----------------------------------------------------------------------------------

Suppose {$b_- \approx b_+$. Then the quantity $-i (b_+-b_-) \omega$ is slowly varying over the range where $-i (b_+ + b_-) \omega$ changes by $2\pi$. 
Let $\omega_*$ be any solution of the approximate QNF condition, and define $b_*= (b_++b_-)/2$. Then for nearby frequencies we are trying to (approximately) solve
 \begin{equation}
\cos(-i\omega_*\pi[b_+ -b_-]) - \cos(-i\omega\pi[b_+ + b_-]) =  2 \cos(\pi\alpha_+)\cos(\pi\alpha_-),
 \end{equation}
 that is
 \begin{equation}
\cos(-i\omega_*\pi[b_+ -b_-]) - \cos(-i\omega2\pi b_*) =  2 \cos(\pi\alpha_+)\cos(\pi\alpha_-),
 \end{equation}
and the solutions of this are approximately
\begin{equation}
\omega_n \approx \omega_* + {in\over b_*}   \qquad    \hbox{valid for} \qquad |n| \ll {b_++b_-\over |b_+-b_-|}.
\end{equation}
Thus approximate result will subsequently be incorporated into a more general perturbative result to be discussed below.

%-----------------------------------------------------------------------------------
\subsection{Case $b_- \ll b_+$}
%-----------------------------------------------------------------------------------

Now suppose {$b_- \ll b_+$. Then the quantity  $-i b_- \omega$ is slowly varying over the range where $-i b_+ \omega$ changes by $2\pi$. 
Let $\omega_*$ be any solution of the approximate QNF condition, then for nearby frequencies we are trying to (approximately) solve
 \begin{equation}
 \sin(-i\omega \pi b_+)\sin(-i\omega_* \pi b_-)  =  \cos(\pi\alpha_+)\cos(\pi\alpha_-),
 \end{equation}
and the solutions of this are approximately
\begin{equation}
\omega_n \approx \omega_* + {2in\over b_+}   \qquad    \hbox{valid for} \qquad |n| \ll {b_+\over b_-}.
\end{equation}

%-----------------------------------------------------------------------------------
\subsection{Case $\alpha_- \approx 1/2$}
%-----------------------------------------------------------------------------------

This corresponds to 
\begin{equation}
V_{0-} b_-^2 \approx 0,
\end{equation}
in which case the QNF condition becomes
 \begin{equation}
 \sin(-i\omega \pi b_+)\sin(-i\omega \pi b_-)  \approx 0.
 \end{equation}
Therefore one obtains either (the physically relevant condition)
 \begin{equation}
 -i\omega b_+=n  \qquad \implies \qquad \omega = {in\over b_+},
 \end{equation}
 or (the physically uninteresting situation)
 \begin{equation}
- i\omega b_-=n.
 \end{equation}
\emph{Note:} If you go to the limit $\alpha_-=1/2$ by setting $V_{0-}=0$, one sees on physical grounds that $b_-$ is irrelevant, so it cannot contribute to the physical QNF. Alternatively if you hold  $V_{0-}\neq0$ but drive $b_-\to 0$, then these QNF's are driven to infinity --- and so decouple from the physics. Either way, the only physically interesting QNFs are $\omega = {in/ b_+}$. We explore these limits more fully below.

%-----------------------------------------------------------------------------------
\section{Some special cases}
%-----------------------------------------------------------------------------------

A number of special cases can now be analyzed in detail to give us an overall feel for the general situation. 

%-----------------------------------------------------------------------------------
\subsection{Case $\alpha_- = 0 = \alpha_+$}
%-----------------------------------------------------------------------------------

This corresponds to 
\begin{equation}
V_{0-} b_-^2 = {1\over 4} = V_{0+} b_+^2,
\end{equation}
in which case the QNF condition becomes
 \begin{equation}
 \sin(i\omega \pi b_+)\sin(i\omega \pi b_-)  = 1.
 \end{equation}
Let us look for pure imaginary QNFs. (We do not  claim that these are the only QNFs.)  This  implies that we must \emph{simultaneously} satisfy \emph{both}
  \begin{equation}
 \sin(-i\omega \pi b_+)= \sin(-i\omega \pi b_-)  = 1,
 \end{equation}
 or \emph{both}
  \begin{equation}
 \sin(-i\omega \pi b_+)= \sin(-i\omega \pi b_-)  = -1.
 \end{equation}
 That is  \emph{both}
\begin{equation}
-i\omega b_+ = 2n_++{1\over2}; \qquad -i\omega b_- = 2n_-+{1\over2},
\end{equation}
or  \emph{both}
\begin{equation}
-i\omega b_+ = 2n_+-{1\over2}; \qquad -i\omega b_- = 2n_--{1\over2}.
\end{equation}
Therefore either
\begin{equation}
{b_+\over b_-} = {2n_++{1\over2}\over 2n_-+{1\over2}}  = {4n_++1\over 4n_-+1}, \qquad \hbox{or} \qquad
{b_+\over b_-} = {2n_+-{1\over2}\over 2n_--{1\over2}} = {4n_+-1\over 4n_--1}.
\end{equation}
In either case we need $b_+/b_-$ to be rational, so that $b_+=p_+ b_*$ and $b_- = p_- b_*$. This special case is thus evidence that there is something very special about the  situation where $b_+/b_-$ is rational, more on this point below. Then either
\begin{equation}
-i\omega = {2n_++{1\over2}\over p_+ b_*} ; \qquad -i\omega ={ 2n_-+{1\over2} \over p_- b_*};
\end{equation}
or
\begin{equation}
-i\omega = {2n_+-{1\over2} \over p_+ b_*} ; \qquad -i\omega  = {2n_--{1\over2} \over p_- b_*}.
\end{equation}
Now write
\begin{equation}
n_+ = m_+ + n p_+; \qquad n_- = m_- + n p_-;
\end{equation}
with $m_+< p_+$ and $m_-<p_-$. (While $n\in\{0,1,2,3,\dots\}$.) Then in the first case
\begin{equation}
\omega = i\left\{ {2m_++{1\over2} \over p_+ b_*}+ {n\over b_*}\right\}  =  i \left\{ {2m_-+{1\over2} \over p_- b_*} + {n\over b_*}\right\}  
=  \omega_* +  {in\over b_*},
\end{equation}
while in the second case
\begin{equation}
\omega = i\left\{ {2m_+-{1\over2} \over p_+ b_*}+ {n\over b_*}\right\}  =  i \left\{ {2m_--{1\over2} \over p_- b_*} + {n\over b_*}\right\}  
=  \omega_* +  {in\over b_*}.
\end{equation}
This is a precursor of the much more general result that we shall ultimately obtain for generic rational $b_+/b_-$.

%-----------------------------------------------------------------------------------
\subsection{Case $V_{0-}=0$}
%-----------------------------------------------------------------------------------

We can best analyze this situation by working directly with the exact wavefunction. 
If  $V_{0-}=0$ then $\alpha_-=1/2$ and
\begin{equation}
\psi_-(0) = 1; \qquad 
\psi'_-(0) =  + i \omega; \qquad
{\psi_-'(0)\over\psi_-(0)}  = + i \omega.
\end{equation}
The exact QNF boundary condition is then
\begin{equation}
i \omega 
= - {2\over b_+} \;  {\Gamma({\alpha_++i\omega b_+\over2} +{3\over4}) \Gamma({-\alpha_++i\omega b_+\over2} +{3\over4}) \over 
\Gamma({\alpha_++i\omega b_+ \over2} + {1\over4}) \Gamma({-\alpha_++i\omega b_+ \over2} + {1\over4})}.
\end{equation}
But this we can rewrite as
\begin{equation}
\omega 
=   {2i \over b_+} \; 
 { \Gamma({-\alpha_+-i\omega b_+ \over2} + {3\over4})  \Gamma({\alpha_+-i\omega b_+ \over2} + {3\over4}) 
 \over
 \Gamma({-\alpha_+-i\omega b_+\over2} +{1\over4})  \Gamma({\alpha_+-i\omega b_+\over2} +{1\over4})} 
 \times 
  {\cos(\pi\alpha_+) - \sin(-i\pi\omega b_+) \over \cos(\pi\alpha_+) + \sin(-i\pi\omega b_+) }.
\end{equation}
This certainly has pure imaginary roots. If we write $\omega = i |\omega|$
then asymptotically ($|\omega|\to\infty$) this becomes
\begin{equation}
1
=   
  {\cos(\pi\alpha_+) - \sin(\pi|\omega| b_+) \over \cos(\pi\alpha_+) + \sin(\pi|\omega| b_+) },
\end{equation}
implying
\begin{equation}
\sin(\pi|\omega| b_+) = 0; \qquad \implies \qquad \pi|\omega| b_+ = n \pi; \qquad \implies \qquad  \omega = {i n\over b_+}.
\end{equation}
This agrees with our previous  calculation for $\alpha_-\approx 1/2$.

%-----------------------------------------------------------------------------------
\subsection{Case $b_{-}\to 0$}
%-----------------------------------------------------------------------------------

This is best dealt with by using a Taylor expansion to show that
\begin{equation}
{\psi_-'(0)\over\psi_-(0)}  = + i \omega + V_{0-} b_- + O(b_-^2) .
\end{equation}
That is
\begin{equation}
\lim_{b_-\to 0} \; {\psi_-'(0)\over\psi_-(0)}  = + i \omega.
\end{equation}
The analysis then follows that for the case $V_{0-}=0$ above, and furthermore agrees with our previous  calculation for $\alpha_-\approx 1/2$.

 %-----------------------------------------------------------------------------------
\subsection{Case $b_{-}\to\infty$}
%-----------------------------------------------------------------------------------
 
This is best dealt with by using the Stirling approximation together with a Taylor expansion to show that
\begin{equation}
{\psi_-'(0)\over\psi_-(0)}  =   i \sqrt{ \omega^2 - V_{0-}} + O(1/b_-^2).
\end{equation}
That is
\begin{equation}
\lim_{b_-\to \infty} \; {\psi_-'(0)\over\psi_-(0)}  = i \sqrt{\omega^2 - V_{0-}}.
\end{equation}
The exact QNF boundary condition is then
\begin{equation}
 i \sqrt{\omega^2 - V_{0-}}
= - {2\over b_+} \;  {\Gamma({\alpha_++i\omega b_+\over2} +{3\over4}) \Gamma({-\alpha_++i\omega b_+\over2} +{3\over4}) \over 
\Gamma({\alpha_++i\omega b_+ \over2} + {1\over4}) \Gamma({-\alpha_++i\omega b_+ \over2} + {1\over4})}.
\end{equation}
But this we can rewrite as
\begin{equation}
\sqrt{\omega^2 - V_{0-}}
=   {2i \over b_+} \; 
 { \Gamma({-\alpha_+-i\omega b_+ \over2} + {3\over4})  \Gamma({\alpha_+-i\omega b_+ \over2} + {3\over4}) 
 \over
 \Gamma({-\alpha_+-i\omega b_+\over2} +{1\over4})  \Gamma({\alpha_+-i\omega b_+\over2} +{1\over4})} 
 \times 
  {\cos(\pi\alpha_+) - \sin(-i\pi\omega b_+) \over \cos(\pi\alpha_+) + \sin(-i\pi\omega b_+) }.
\end{equation}
If we write $\omega = i |\omega|$
then asymptotically, ($|\omega|\to\infty$, with $V_{0-}$ held fixed, implying that $V_{0-}$ effectively decouples from the calculation), this becomes
\begin{equation}
1
=   
  {\cos(\pi\alpha_+) - \sin(\pi|\omega| b_+) \over \cos(\pi\alpha_+) + \sin(\pi|\omega| b_+) },
\end{equation}
implying
\begin{equation}
\sin(\pi|\omega| b_+) = 0; \qquad \implies \qquad \pi|\omega| b_+ = n \pi; \qquad \implies \qquad  \omega = {i n\over b_+}.
\end{equation}
The importance of this observation is that it indicates that for ``one sided'' potentials it is only the side for which the potential has exponential falloff that contributes to the ``gap''.

%----------------------------------------------------------------------------------------------------------------------------------------
\section{Rational and irrational ratios for the falloff}\label{S:rational}
%----------------------------------------------------------------------------------------------------------------------------------------

We have already seen above good reason to suspect that rational ratios $b_+/b_-$ might be special. Let us now explore this case in more detail. 

%-----------------------------------------------------------------------------------
\subsection{Explicit examples}
%-----------------------------------------------------------------------------------

\begin{itemize}

\item If $b_+=b_- = b_*$, but we do not necessarily demand $\alpha_+=\alpha_-$, then the asymptotic QNFs are exactly calculable and are given by
\begin{equation}
\omega_n  =  {i\cos^{-1}\left\{1 - 2  \;  \cos(\pi\alpha_+)\cos(\pi\alpha_-)\right\}\over2\pi b_*} + {in\over b_*}.
\end{equation} 

\emph{Proof:} 
The asymptotic QNF condition reduces to
\begin{equation}
1 - \cos(-i2\pi\omega b_*) = 2  \;  \cos(\pi\alpha_+)\cos(\pi\alpha_-),
 \end{equation}
whence
\begin{equation}
 \cos(-i2\pi\omega b_*) =  1 - 2  \;  \cos(\pi\alpha_+)\cos(\pi\alpha_-).
 \end{equation}
This is easily solved to yield
\begin{equation}
 -i2\pi\omega_n b_* =  \cos^{-1}\left\{1 - 2  \;  \cos(\pi\alpha_+)\cos(\pi\alpha_-)\right\} + 2n\pi,
 \end{equation}
 whence
\begin{equation}
\omega_n  =  {i\cos^{-1}\left\{1 - 2  \;  \cos(\pi\alpha_+)\cos(\pi\alpha_-)\right\}\over2\pi b_*} + {in\over b_*}.
 \end{equation} 

\emph{Comment:} These QNFs are pure imaginary for $\cos(\pi\alpha_+)\cos(\pi\alpha_-)\leq 1$, and off-axis complex for  $\cos(\pi\alpha_+)\cos(\pi\alpha_-) > 1$. 
We can always, for convenience, choose to define $\Re(\cos^{-1}(x)) \in [0,2\pi)$; then $\cos^{-1}(\cdot)$ is double valued.
\begin{equation}
\theta\in[0,\pi] \hbox{ and } \cos\theta=x \hbox{ implies }\cos^{-1}(x) = \{\theta, \pi-\theta\}, 
\end{equation}
\begin{equation}
\theta\in[\pi,2\pi) \hbox{ and } \cos\theta=x \hbox{ implies } \cos^{-1}(x) = \{\theta, 3\pi-\theta\} . 
\end{equation}
With this notation
\begin{equation}
\omega_n  =    \omega_0 + {in\over b_*}; \qquad   \omega_0 =  {i\cos^{-1}\left\{1 - 2  \;  \cos(\pi\alpha_+)\cos(\pi\alpha_-)\right\}\over2\pi b_*};
 \end{equation} 
 with $0\leq \Im(\omega_0) < 1/b_*$ and $n\in\{0,1,2,3,\dots\}$. Note that because of the double valued nature of  $\cos^{-1}(\cdot)$ there are actually two branches of QNFs hiding in this notation --- which we will need if we wish to regain the known standard result when we  specialize to $\alpha_-=\alpha_+$. (We shall subsequently generalize this specific result, but it is explicit enough and compact enough to make it worthwhile presenting it in full. Furthermore we shall need this as input to our perturbative analysis.)

\item If $b_+=3b_-$, that is $b_+={3\over2} b_*$ and $b_-={1\over2} b_*$, but we do not necessarily demand $\alpha_+=\alpha_-$, then the asymptotic QNFs are calculable and are given by
\begin{equation}
\omega_n  =  {i\over \pi b_*} \cos^{-1}\left(  {1 \pm \sqrt{9 - 16  \;  \cos(\pi\alpha_+)\cos(\pi\alpha_-))} \over 4} \right) + {2in\over b_*}.
\end{equation} 

\emph{Proof:}  To see this note that in this situation $|b_+-b_-| = b_* = (b_++b_-)/2$. Therefore the QNF condition reduces to
\begin{equation}
 \cos(-i\pi\omega b_*) - \cos(-i2\pi\omega b_*) = 2  \;  \cos(\pi\alpha_+)\cos(\pi\alpha_-),
\end{equation}
implying
\begin{equation}
 \cos(-i\pi\omega b_*) - 2\cos^2(-i\pi\omega b_*) + 1 = 2  \;  \cos(\pi\alpha_+)\cos(\pi\alpha_-).
\end{equation}
That is
\begin{equation}
2\cos^2(-i\pi\omega b_*) -  \cos(-i\pi\omega b_*)  - 1 +  2  \;  \cos(\pi\alpha_+)\cos(\pi\alpha_-) = 0,
\end{equation}
whence
\begin{equation}
\cos(-i\pi\omega b_*)  = {1 \pm \sqrt{ 1 + 8(1-2  \;  \cos(\pi\alpha_+)\cos(\pi\alpha_-))} \over 4},
\end{equation}
so that
\begin{equation}
\cos(-i\pi\omega b_*)  = {1 \pm \sqrt{9 - 16  \;  \cos(\pi\alpha_+)\cos(\pi\alpha_-))} \over 4},
\end{equation}
implying
\begin{equation}
-i\pi\omega_n b_* = \cos^{-1}\left(  {1 \pm \sqrt{9 - 16  \;  \cos(\pi\alpha_+)\cos(\pi\alpha_-))} \over 4} \right) + n 2\pi.
\end{equation}
Finally
\begin{equation}
\omega_n  =  {i\over \pi b_*} \cos^{-1}\left(  {1 \pm \sqrt{9 - 16  \;  \cos(\pi\alpha_+)\cos(\pi\alpha_-))} \over 4} \right) + {2in\over b_*}.
\end{equation} 
\emph{Comment:} This gives us another specific example of asymptotic off-axis complex QNF's --- now with $b_+\neq b_-$. Note that because of the $\pm$ and the double-valued nature of $\cos^{-1}(\cdot)$ there are actually 4 branches of QNFs hiding in this notation.

\item This particular trick can certainly be extended to the cubic and quartic polynomials, for which general solutions exist.
\begin{itemize}
\item 
The quadratic corresponds to
\begin{equation}
{b_+-b_-\over b_++b_-} = 2; \qquad b_+ = 3 \;b_-.
\end{equation}
\item
The cubic corresponds to
\begin{equation}
{b_+-b_-\over b_++b_-} = 3; \qquad b_+ = 2 \; b_-.
\end{equation}
\item
The quartic corresponds to 
\begin{equation}
{b_+-b_-\over b_++b_-} = 4; \qquad b_+ = {5\over3}\; b_-.
\end{equation}
\end{itemize}
This very strongly suggests there is something special about arbitrary rational values of $b_+/b_-$.
In fact the current results will very shortly be subsumed into much more general results for rational values of $b_+/b_-$.

\end{itemize}

%-----------------------------------------------------------------------------------
\subsection{Rational ratios for the falloff}
%-----------------------------------------------------------------------------------

Suppose $b_+/b_-$ is rational, that is
\begin{equation}
{b_+\over b_-} = {p_+\over p_-},
\end{equation}
and suppose we now define $b_*$ by
\begin{equation}
b_+= p_+ b_*; \qquad b_- = p_- b_*; \qquad b_* = \mathrm{hcf}(b_+, b_-),
\end{equation}
then the asymptotic QNF condition is given by
 \begin{equation}
 \sin(-i\omega \pi p_+ b_*)\sin(-i\omega \pi p_- b_*)  =  \cos(\pi\alpha_+)\cos(\pi\alpha_-).
 \end{equation}
If $\omega_*$ is \emph{any} specific solution of this equation, then (setting $s=1$ for $p_+ p_-$ odd, and $s=2$ for $p_+ p_-$ even)
\begin{equation}
\omega_n = \omega_* + {ins\over b_*} = \omega_* + ins \; \mathrm{lcm}\left({1\over b_+}, {1\over b_-}\right)
\end{equation}
will also be a solution.
But are these the only solutions? Most definitely not.
For instance, consider (for rational $b_+/b_-$) the set of all QNFs for which 
\begin{equation}
\Im(\omega) < {s\over b_*},
\end{equation}
and label them as
\begin{equation}
\omega_{0,a} \qquad a\in\{1,2,3\dots N\}.
\end{equation}
Then the set of all QNFs decomposes into a set of families
\begin{equation}
\omega_{n,a} = \omega_{0,a} + {ins\over b_*}; \qquad a\in\{1,2,3\dots N\}; \qquad n\in\{0,1,2,3\dots\};
\end{equation}
where $N$ is yet to be determined. 
But for rational  $b_+/b_-$  we can rewrite the QNF condition as
\begin{equation}
\cos(-i\omega\pi b_*|p_+-p_-|) - \cos(-i\omega\pi b_*[p_+ +p_-])  = 2\cos\left(\pi\alpha_+\right) \cos\left(\pi\alpha_-\right).
\end{equation}
Now define $z=\exp(\omega\pi b_*)$,  then the QNF condition
can be rewritten as
\begin{equation}
z^{|p_+-p_-|} + z^{-|p_+-p_-|} - z^{[p_+ +p_-]} - z^{-[p_+ +p_-]} =  4\cos\left(\pi\alpha_+\right) \cos\left(\pi\alpha_-\right),
\end{equation}
or equivalently
\begin{equation}
z^{2[p_+ +p_-]} - z^{|p_+-p_-|+[p_+ +p_-]} + 4\cos\left(\pi\alpha_+\right) \cos\left(\pi\alpha_-\right) z^{+[p_+ +p_-]} -  z^{-|p_+-p_-|+[p_+ +p_-]} + 1 = 0,
\end{equation}
that is
\begin{equation}
z^{2[p_+ +p_-]} - z^{2p_\mathrm{max}} + 4\cos\left(\pi\alpha_+\right) \cos\left(\pi\alpha_-\right) z^{+[p_+ +p_-]} -  z^{2p_\mathrm{min}} + 1 = 0.
\end{equation}
This is a polynomial of degree $N=2(p_++p_-)$, so it has exactly $N$ roots $z_a$ (occurring in complex conjugate pairs, and whenever  $p_+ p_-$ is odd, in symmetric pairs $\pm z_a$). Then the QNFs are, with the imaginary part of the logarithm lying in $[0,2\pi)$,
 \begin{equation}
 \omega_{n,a} = 
  { \ln(z_a) \over \pi b_*} + {ins\over b_*}
 \qquad  a\in\{1,2,3\dots s(p_++p_-)\} \qquad n\in\{0,1,2,3,\dots\}. 
 \end{equation}
 So for rational $b_+/b_-$ with $b_+/b_- = p_+/p_-$ we have exactly $s(p_++p_-)$ equi-spaced families of QNFs all with with gap $is/b_*$ and with (typically distinct) offsets $  { \ln(z_a) /( \pi b_*) } $. 
 That is: \emph{Arbitrary rational ratios of ${b_+/ b_-} $ automatically  imply the}  $\omega_n =  \hbox{(offset)} + i n \; \hbox{(gap)}$ \emph{behaviour.}

%-----------------------------------------------------------------------------------
\paragraph{Comparison with some monodromy results: }
%-----------------------------------------------------------------------------------
Consider the specific situation presented in~\cite{Natario}, that is, Schwarzschild-de Sitter (Kottler) spacetime with surface gravity $k_H$ at the black hole horizon and $k_C$ at the cosmological horizon.  In that article monodromy techniques are used to obtain an approximate QNF condition (for scalar fields) of the form
\begin{equation}
\cosh\left({\pi\omega\over k_H}- {\pi\omega\over k_C}\right) + 3 \cosh\left({\pi\omega\over k_H}+ {\pi\omega\over k_C}\right) = 0.
\end{equation}
Qualitatively similar QNF conditions hold for higher spin.

Now if $k_H/k_C = n_H/n_C$ is rational then we can (strategically) choose integers $n_C$ and $n_H$ so that $k_H/k_C = n_C/n_H$, and so write $k_H = k_*/n_H$ while $k_C =  k_*/n_C$.  Then this approximate QNF condition becomes
\begin{equation}
\cosh\left({n_H\pi\omega\over k_*}- {n_C\pi\omega\over k_*}\right) + 3 \cosh\left({n_H\pi\omega\over k_*}+ {n_C\pi\omega\over k_*}\right) = 0.
\end{equation}
Defining $z= \exp(\pi \omega/k_*)$ this becomes the (Laurent) polynomial condition
\begin{equation}
z^{n_H-n_C}  + z^{n_C-n_H} + 3 z^{n_H+n_C} + 3 z^{-n_H-n_C} = 0,
\end{equation}
which we can rearrange as the (ordinary) polynomial
\begin{equation}
3 z^{2(n_H+n_C)} + z^{2n_H}  + z^{2n_C}  + 3 = 0.
\end{equation}
This is an ordinary polynomial with degree $N= 2(n_H+n_C)$ so it has $N= 2(n_H+n_C)$  roots $z_a$ (typically distinct, apart from the fact that they occur in complex conjugate and symmetric pairs $\pm z_a$).  Thus in a manner qualitatively similar to our direct semi-analytic results above, the monodromy based approximate QNFs will fall into families of the form
 \begin{equation}
 \omega_{n,a} = 
  { \ln(z_a) \over \pi k_*} + {in\over k_*}
 \qquad  a\in\{1,2,3\dots (n_H+n_C)\} \qquad n\in\{0,1,2,3,\dots\}. 
 \end{equation}
In each individual family the real part of the QNF does not oscillate. As one moves from one family to the next the real part of the QNF will change. If the degree of the polynomial $N=2(n_H+n_C)$ is sufficiently high then the discrete ``base frequencies'' $ { \ln(z_a) / \pi k_*}$ will visually approximate a single cycle of an oscillating curve. In figure 3 of ref~\cite{Natario} several thousand families of equi-spaced QNFs combine to yield a discrete collection of points that visually appears to be a repetitive oscillating curve.  That is,  the present semi-analytic techniques, where they overlap, are compatible with expectations based on monodromy techniques.

%-----------------------------------------------------------------------------------
\subsection{Irrational ratios for the falloff}
%-----------------------------------------------------------------------------------

Now suppose $b_+/b_-$ is irrational, that is
\begin{equation}
b_*=  \mathrm{hcf}(b_+, b_-) = 0.
\end{equation}
Then all of the ``families'' considered above only have one element
\begin{equation}
\omega_{0,a}  \qquad a\in\{1,2,3\dots \infty\}. 
\end{equation}
That is, there will be no ``pattern'' in the QNFs, and they will not be regularly spaced. (Conversely, if there is a ``pattern'' then  $b_+/b_-$ is rational.)
Stated more formally, it is possible to derive a theorem as below.

\bigskip

\noindent
{\bf Theorem:}
Suppose we have at least one family of equi-spaced QNFs such that
\begin{equation}
\label{E:equal}
\omega_n = \omega_0 + i n K,
\end{equation}
then  $b_+/b_-$ is rational.

\bigskip

\noindent
{\bf Proof:} If we have a family of QNFs of the form given in equation (\ref{E:equal})
then we know that $\forall n \geq 0$
\begin{eqnarray}
&&
\cos(-i\omega_0\pi |b_+-b_-| + nK\pi |b_+-b_-| ) - \cos(-i\omega_0\pi [b_+ +b_-] + nK\pi |b_++b_-| ) \qquad
\nonumber\\
&& \qquad\qquad \qquad 
= \cos(-i\omega_0\pi |b_+-b_-| ) - \cos(-i\omega_0\pi [b_+ +b_-]  ).
\end{eqnarray}
Let us write this in the form $\forall n \geq 0$
\begin{equation}
\cos(A+n J) - \cos(B+nL) = \cos(A)-\cos(B),
\end{equation}
and realize that this also implies
\begin{equation}
\cos(A+[n+1] J) - \cos(B+[n+1]L) = \cos(A)-\cos(B),
\end{equation}
and
\begin{equation}
\cos(A+[n+2] J) - \cos(B+[n+2]L) = \cos(A)-\cos(B). 
\end{equation}
Now appeal to the trigonometric identity (based on equation (\ref{E:trig3}))
\begin{equation}
\cos(A+[n+2] J) + \cos(A+n J) = 2 \cos(J) \cos(A+[n+1] J),
\end{equation}
to deduce
\begin{equation}
\cos(J) \cos(A+[n+1] J) - \cos(L) \cos(B+[n+1]L) = \cos(A)-\cos(B).
\end{equation}
That is, $\forall n\geq 0$ we have \emph{both}
\begin{equation}
\cos(A+[n+1] J) - \cos(B+[n+1]L) = \cos(A)-\cos(B),
\end{equation}
\emph{and}
\begin{equation}
\cos(J) \cos(A+[n+1] J) - \cos(L) \cos(B+[n+1]L) = \cos(A)-\cos(B).
\end{equation}
The first of these equations asserts that all the points 
\begin{equation}
\left( \vphantom{\Big|} \cos(A+[n+1] J) , \; \cos(B+[n+1]L)  \right)
\end{equation} 
lie on the straight line of slope 1 that passes through the point $(0, \cos B-\cos A)$. 
The second of these equations asserts that all the points 
\begin{equation}
\left( \vphantom{\Big|}  \cos(A+[n+1] J) , \; \cos(B+[n+1]L)  \right)
\end{equation}
\emph{also} lie on the straight line of slope $\cos(J)/\cos(L)$ that passes through the point $(0, [\cos B-\cos A]/\cos L)$. We then argue as follows:
\begin{itemize}
\item If $\cos J \neq \cos L$ then these two lines are not parallel and so meet only at a single point, let's call it $(\cos A_*, \cos B_*)$, whence we deduce
\begin{equation}
\cos(A+[n+1] J) = \cos A_* ; \qquad \cos(B+[n+1]L) = \cos B_*.
\end{equation}
But then both $J$ and $L$ must be multiples of $2\pi$, and so $\cos J = 1=  \cos L$ contrary to hypothesis.
\item
If  $\cos J = \cos L \neq 1$ then we have both
\begin{equation}
\cos(A+[n+1] J) - \cos(B+[n+1]L) = \cos(A)-\cos(B),
\end{equation}
and
\begin{equation}
\cos(J)\left[  \cos(A+[n+1] J) -  \cos(B+[n+1]L) \right] = \cos(A)-\cos(B).
\end{equation}
but these are two parallel lines, both of slope 1, that never intersect unless $\cos(J)=1$.
Thus  $\cos J = 1=  \cos L$ contrary to hypothesis.
\item
We therefore conclude that both $J$ and $L$ must be multiples of $2\pi$ so that $\cos J = 1=  \cos L$ (in which case the QNF condition is certainly satisfied).
\end{itemize}
But now
\begin{equation}
{|b_+-b_-|\over b_++b_-} = {J\over L} \; \in\; Q,
\end{equation}
and therefore 
\begin{equation}
{b_+\over b_-}  \;\in\; Q.
\end{equation}
That is: \emph{Rational ratios of ${b_+/ b_-} $ are implied by the}  $\omega_n =  \hbox{(offset)} + i n \; \hbox{(gap)}$ \emph{behaviour}.

%----------------------------------------------------------------------------------------------------------------------------------------
\section{Systematic first-order perturbation theory}\label{S:perturbation}
%----------------------------------------------------------------------------------------------------------------------------------------

Sometimes it is worthwhile to adopt a perturbative approach and to estimate shifts in the QNFs from some idealized pattern. 
Define
\begin{equation}
b ={ b_++b_-\over2};  \qquad  \Delta = |b_+-b_-|;
\end{equation}
and rewrite the asymptotic QNF condition as
\begin{equation}
 \cos(-i\pi\omega\Delta) - \cos(-i2\pi\omega b) = 2  \;  \cos(\pi\alpha_+)\cos(\pi\alpha_-),
 \end{equation}
where we are implicitly holding $\alpha_\pm$ fixed.
When $\Delta=0$ we have previously seen that  the QNF are explicitly calculable with 
\begin{equation}
\hat \omega_n  =  {i\cos^{-1}\left\{1 - 2  \;  \cos(\pi\alpha_+)\cos(\pi\alpha_-)\right\}\over2\pi b} + {in\over b}.
 \end{equation} 
Can we now obtain an approximate formula for the  the QNF's when $\Delta\neq 0$?  It is a good strategy to define the dimensionless parameter $\epsilon$ by 
\begin{equation}
\Delta = 2\; \epsilon \; b,
\end{equation} 
and to set
\begin{equation}
\omega= \hat \omega + \delta\omega; \qquad   \delta\omega = O(\epsilon);
\end{equation}
so that the asymptotic QNF condition becomes
\begin{equation}
\cos(-i2\pi[\hat \omega +\delta\omega]\epsilon b) - \cos(-i2\pi[\hat \omega+\delta\omega] b) = 2  \;  \cos(\pi\alpha_+)\cos(\pi\alpha_-).
\end{equation}
Then to first order in $\epsilon$
\begin{equation}
\cos(-i2\pi\hat\omega \epsilon b) - \cos(-i2\pi[\hat\omega+\delta\omega] b) = 2  \;  \cos(\pi\alpha_+)\cos(\pi\alpha_-),
\end{equation}
where implicitly this approximation requires $\epsilon |\delta\omega|  b \ll 1$. 
Subject to this condition we have
\begin{equation}
\cos(-i2\pi[\hat \omega+\delta\omega] b) = \cos(-i2\hat \pi\omega \epsilon b) - 2  \;  \cos(\pi\alpha_+)\cos(\pi\alpha_-),
\end{equation}
whence
\begin{equation}
-i2\pi[\hat \omega_n+\delta\omega_n] b = \cos^{-1} \left\{ \cos(-i2\pi\hat\omega_n \epsilon b) - 2  \;  \cos(\pi\alpha_+)\cos(\pi\alpha_-) \right\} + 2\pi n,
\end{equation}
so that
\begin{equation}
\hat \omega_n+\delta\omega_n  = i {\cos^{-1} \left\{ \cos(-i2\pi\hat \omega_n \epsilon b) - 2  \;  \cos(\pi\alpha_+)\cos(\pi\alpha_-) \right\}\over 2\pi b} + {in\over b}.
\end{equation}
But we know that the unperturbed QNFs satisfy $\hat\omega_n = \hat\omega_0 +{in/b}$, so we can also write this as
\begin{equation}
\delta\omega_n  = i {\cos^{-1} \left\{ \cos(-i2\pi\hat \omega_n \epsilon b) - 2  \;  \cos(\pi\alpha_+)\cos(\pi\alpha_-) \right\}\over 2\pi b} - \hat\omega_0.
\end{equation}
Using the definition of $\hat\omega_n$ this can now be cast in the form
\begin{equation}
\delta\omega_n  = i {\cos^{-1} \left\{ \cos(-i2\pi\hat \omega_n \epsilon b)  +  \cos(-i2\pi\hat \omega_n  b)  - 1 \right\}\over 2\pi b} - \hat\omega_0,
\end{equation}
or the slightly more suggestive
\begin{equation}
\delta\omega_n  = i {\cos^{-1} \left\{ \cos(-i2\pi\hat \omega_n b)  +  \cos(-i2\pi\hat \omega_n  \epsilon b)  - 1 \right\}\over 2\pi b} - \hat\omega_0,
\end{equation}
which can even be simplified to
\begin{equation}
\delta\omega_n  = i \; {\cos^{-1} \left\{ \cos(-i2\pi\hat \omega_0 b)  +  \cos(-i2\pi\hat \omega_n  \epsilon b)  - 1 \right\}\over 2\pi b} - \hat\omega_0.
\end{equation}
Note that this manifestly has the correct limit as $\epsilon\to0$. 
Note that we have \emph{not} asserted or required that $\hat\omega_n \, \epsilon \, b \ll 1$, in fact when $n \gg 1/\epsilon$ this is typically \emph{not} true.
(Consequently $\cos(-i2\pi\hat\omega_n \epsilon b) $ is relatively unconstrained.)
Note furthermore that $\Im(\delta\omega_n) \leq 1/b$.

%----------------------------------------------------------------------------------------------------------------------------------------
\section{Discussion}\label{S:discussion}
%----------------------------------------------------------------------------------------------------------------------------------------

The key lesson to be learned from our semi-analytic model for the QNFs is that the  $\omega_n =  \hbox{(offset)} + i n \; \hbox{(gap)}$ behaviour  is \emph{common but not universal}.  Specifically, in our semi-analytic model the key point is whether or not the ratio ${b_+/ b_-} $ is a rational number.  Let us also note here that monodromy techniques (see in particular~\cite{Natario}) quite often also lead to the qualitative result that rational ratios of surface gravities are closely related to the occurence of $\omega_n =  \hbox{(offset)} + i n \; \hbox{(gap)}$ behaviour.  This behaviour is thus ``common but not universal'' in the sense that mathematically the rational numbers are dense in the reals, while when applying numerical techniques floating point numbers are a subset of the rational numbers. So the  $\omega_n =  \hbox{(offset)} + i n \; \hbox{(gap)}$ behaviour is actually universal within the context of floating point arithmetic. 

Will this behaviour extend to more ``realistic'' astrophysical or asymptotically de~Sitter black holes? Consider a ``wavepacket'', built up out of  highly damped modes, that is  centered near the peak of the Regge--Wheeler (Zerelli) potential. While the initial short-time behaviour of the wavepacket is likely to be sensitive to the details of the Regge--Wheeler (Zerelli) potential, such a wavepacket will quickly damp out and spread out towards both $r_*\to-\infty$ and $r_*\to +\infty$, so that the wavepacket will penetrate regions where our piecewise Eckart model potential, (governed by the surface gravities at the event and cosmological horizons), should be a good approximation to the true potential. We should therefore expect the results of our semi-analytic model to be qualitatively (but not necessarily quantitatively) accurate for estimating the asymptotic QNFs of ``realistic'' asymptotically de~Sitter black holes. Because of the way the asymptotic QNF condition was derived, we do not expect out model to give good results for low-lying QNFs. 

Overall, one of the nice features of this semi-analytic model is that a quite surprising amount of semi-analytic information can be extracted,  in terms of general qualitative results,  approximate results,  perturbative results, and reasonably explicit computations. We suspect that it might be possible to generalize the model potential even further --- the ``art'' would lie in picking a piecewise potential that is still analytically solvable (at least for the highly damped modes) but which might be closer in spirit to the Regge--Wheeler (Zerelli) potential that is the key physical motivation for the current article. (Of course if we temporarily forget the black hole motivation, it may already be of some mathematical and physical interest that we have a nontrivial extension of the Eckart potential that is asymptotically exactly solvable --- one could in principle loop back to Eckart's original article and start asking questions about tunelling probabilities for electrons encountering such piecewise Eckart barriers.)

%----------------------------------------------------------------------------------------------------------------------------------------
%\clearpage
%----------------------------------------------------------------------------------------------------------------------------------------
\appendix
%-----------------------------------------------------------------------------------------------------------------------------------------
\section{Trigonometric identities}\label{A:trig}
%-----------------------------------------------------------------------------------------------------------------------------------------

In the body of the article we will need to use some slightly unusual trigonometric identities. They can be derived from standard ones without too much difficulty but  are sufficiently unusual to be worth mentioning explicitly:
 \begin{equation}
 \label{E:trig1}
 \tan A \; \tan B = { \cos(A-B) - \cos(A+B)\over\cos(A-B) + \cos(A+B)};
 \end{equation}
 \begin{equation}
  \label{E:trig2}
 \tan\left({ A+B\over2}\right)  \; \tan\left({ A-B\over2}\right) = { \cos B- \cos A\over\cos B + \cos A};
 \end{equation}
 and
 \begin{equation}
  \label{E:trig3}
\cos(A+2B) + \cos A = 2 \cos B \; \cos(A+B).
\end{equation}

%-----------------------------------------------------------------------------------------------------------------------------------------
\section{Gamma function results}\label{A:gamma}
%-----------------------------------------------------------------------------------------------------------------------------------------
The key Gamma function identity we need is
\begin{equation}
\label{E:reflection}
\Gamma(z)\; \Gamma(1-z) = {\pi\over\sin(\pi z)}.
\end{equation}
We also need the following asymptotic estimate based on the Stirling approximation
\begin{equation}
\label{E:stirling}
 {\Gamma(z+{1\over2})\over\Gamma(z)} = \sqrt{z} \;  \left[1+ O\left({1\over z}\right)\right];   \qquad\qquad \Re(z)\to \infty.
\end{equation}

%-----------------------------------------------------------------------------------------------------------------------------------------
\section{Hypergeometric function identities}\label{A:hyper}
%-----------------------------------------------------------------------------------------------------------------------------------------
The key hypergeometric function identities we need are Bailey's theorem
\begin{equation}
\label{E:bailey}
_2F_1\left(a,1-a,c,{1\over2}\right) = {\Gamma({c\over2}) \Gamma({c+1\over2}) \over \Gamma({c+a\over2}) \Gamma({c-a+1\over2})},
\end{equation}
which is easily found in many standard references, 
and the particular differential identity
\begin{equation}
\label{E:differential}
{\d \left\{ _2F_1\left(a,b,c,z\right) \right\} \over \d z} = {c-1\over z} \left[  \; _2F_1\left(a,b,c-1,z\right) -  \;_2F_1\left(a,b,c,z\right) \right],
\end{equation}
which is not found in any of the standard references (but is easy enough to verify once it has been presented).

%----------------------------------------------------------------------------------------------------------------------------------------
%\clearpage
%----------------------------------------------------------------------------------------------------------------------------------------

%---------------------------------------------------------------------------------------------------------------------------------------

%------------------------------------------------------------------------------------------------------------------------------------------
\end{document}